\documentclass[a4paper,12pt]{JHEP3} 

\usepackage[all]{xy} 

\newcommand{\del}{\partial}   
   
\def\drv#1{{\partial_{#1}}}   
\def\drvstar#1{\partial\kern-0.5pt\smash{\raise 4.5pt\hbox{$\ast$}}  
               \kern-5.0pt_{#1}}   
\def\lvec#1{\setbox0=\hbox{$#1$}  
    \setbox1=\hbox{$\scriptstyle\leftarrow$}  
    #1\kern-\wd0\smash{  
    \raise\ht0\hbox{$\raise1pt\hbox{$\scriptstyle\leftarrow$}$}}  
    \kern-\wd1\kern\wd0}   
   
\def\ldrvstar#1{\lvec{\,\partial}\kern-0.5pt\smash{\raise 4.5pt\hbox{$\ast$}}  
               \kern-5.0pt_{#1}}   
\title{Perturbative dynamics of matrix string for the membrane}  
\author{ 
 Masashi Hayakawa \\ 
 Theoretical Physics Group, RIKEN, 
 Wako 2-1, Saitama 351-0198, Japan \\ 
 \email{haya@riken.jp} 
} 
\author{ 
 Nobuyuki Ishibashi \\ 
 Institute of Physics, University of Tsukuba, 
 Tsukuba, Ibaraki 305-8571, Japan \\ 
 \email{ishibash@het.ph.tsukuba.ac.jp} 
} 

\abstract{ 
 Recently Sekino and Yoneya proposed a way 
to regularize the world volume theory of membranes 
wrapped around $S^1$ by matrices 
and showed that one obtains matrix string theory 
as a regularization of such a theory.   
 We show that this correspondence between matrix string theory   
and wrapped membranes can be obtained 
by using the usual M(atrix) theory techniques.   
 Using this correspondence, 
we construct the super-Poincar\'e generators 
of matrix string theory at the leading order in the   
perturbation theory.   
 It is shown 
that these generators satisfy 
10 dimensional super-Poincar\'e algebra without any anomaly. 
}  
  
\preprint{ 
 {\bf January 2004} \\ 
 {\bf UTHEP-483} \\ 
 {\bf hep-th/0401227} 
} 
  

\begin{document} 
\section{Introduction}   
\label{sec:intro}   
 In order to define the world volume theory of membranes,   
one needs a way to regularize the ultraviolet divergences.   
 The matrix regularization is the most useful way to do so   
after the light-cone gauge fixing of the world volume diffeomorphism   
\cite{matrix_reg,deWit:1988ig} .   
 The dependence on the two spatial world volume coordinates   
are represented by matrices,   
and the regularized theory becomes a quantum mechanical system   
described by a set of matrix variables.   
 The area preserving diffeomorphism, 
which is the residual symmetry in the light-cone gauge,   
is replaced by $U(N)$ gauge symmetry   
and the resulting theory becomes     
a one-dimensional gauge system.   
 Later it was shown that the same system can be obtained   
from string theory, trying to describe the DLCQ 
(discretized light-cone quantization) limit   
of M theory \cite{Banks:1996vh,Sen:1997we,Seiberg:1997ad}.   
   
 The above matrix quantum mechanics provides a regularization of   
{\it unwrapped} membranes.   
 In M theory \cite{M_theory} ,   
one is tempted to consider {\it wrapped} membranes,   
i.e., membranes with one spatial world volume direction   
wrapped around a compactified direction in the space-time,   
which become fundamental strings   
when the compactification radius is taken to be very small. 
 Wrapped membranes were first studied in detail 
in Ref.~\cite{deWit:1997zq}. 
 Recently, Sekino and Yoneya gave a recipe \cite{Sekino:2001ai}   
to regularize the world volume theory of wrapped membranes   
by using matrices.   
 They showed that by applying their rules one obtains matrix string   
theory \cite{matrix_string} as a regularization of the world volume theory   
of wrapped membranes. 
(See Ref.~\cite{Uehara:2002ga,Cederwall:2002wh,Shimada:2003ks} 
for related works, 
and Ref.~\cite{Recent_membrane} 
for recent other approaches to the membrane theory.). 
   
  Matrix string theory was originally obtained   
from M(atrix) theory \cite{Banks:1996vh}   
via the procedure in Ref.~\cite{Taylor:1996ik},   
in order to describe M theory with one space direction compactified.   
(See the appendix of Ref.~\cite{Giddings:1998yd} for details.).   
 Since it was derived following the logic of string theory and M theory,   
it is likely that the matrix regularization of Sekino and Yoneya   
can also be rederived from string theory and M theory as in the   
unwrapped case.   
 In the first half of this paper,   
we would like to show that it is indeed the case and   
the matrix regularization rule of Sekino and Yoneya can be rederived   
using the familiar techniques in M(atrix) theory.   
  
 Therefore, 
matrix string theory (or a two dimensional super Yang-Mills theory)   
can be considered as a microscopic    
description of the world volume theory 
of wrapped membranes in M theory.   
 In the Sekino-Yoneya regularization, 
each matrix element of 
the matrix variables 
corresponds to a tiny string bit on the world volume of the membrane.   
 Therefore in order to get information 
on the world volume theory of the wrapped membranes, 
we need to analyze the infrared properties of 
the two dimensional $U(N)$ super Yang-Mills theory.   
 Here, we would like to concentrate on the ultraviolet   
region where we can calculate various quantities perturbatively.   
 The quantities which we are particularly interested in are the   
anomalies in the super-Poincar\'e algebra.   
 Requiring that such anomalies vanish, we can determine the   
critical dimension   
of matrix string theory.   
  
 In the latter half of this paper, 
we will construct the super-Poincar\'e generators 
of matrix string theory using its relation with wrapped membranes.   
 We see that 
at the leading order of the perturbative expansion    
the theory can be described as a theory of $N$ strings   
without any interactions between them.   
 At this order,   
the membrane appears to be resolved into the multi-strings put   
at the sites along the compact `11th' direction \cite{HI}.   
 As a consequence,   
the super-Poincar\'e generators are the sum of those for these strings   
and they satisfy the super-Poincar\'e algebra 
when the number of space-time dimensions 
is equal to $10$ for the strings. 
 Since the membrane is wrapped around one direction, 
the membrane lies in $11$ dimensional space-time.  
  
 The organization of this paper is as follows.   
 In Sec.~\ref{sec:membrane_matrix},   
we briefly review the matrix regularization 
of wrapped membrane action proposed 
by Sekino and Yoneya.   
 In Sec.~\ref{sec:d1},   
we show that the rule given by Sekino and Yoneya can be derived   
using the usual M(atrix) theory techniques.   
 In Sec.~\ref{sec:Lorentz}, we will examine 
the super-Poincar\'e invariance of matrix string theory using   
its relation to the wrapped membrane. 
 We will show that 
the super-Poincar\'e generators of the theory 
can be written mostly 
as the sum of those of $N$ independent Type IIA strings 
and that there exists super-Poincar\'e symmetry 
at the lowest order in perturbation theory. 
 Sec.~\ref{sec:concl} is devoted to 
conclusions and discussions.   
  
\section{Matrix regularization of wrapped membranes}   
\label{sec:membrane_matrix}   
 We recall briefly    
the direct construction of the map of   
the membrane variables to the matrices in matrix string theory  
\'a la Sekino and Yoneya \cite{Sekino:2001ai}   
after fixing the notation used throughout this paper.   
  
\subsection{light-cone gauge fixing}   
\label{sec:lc}   
  
 Here we recall the conventional light-cone gauge action 
for the supermembrane.   
 The Nambu-Goto action for the supermembrane propagating   
in 11-dimensional Minkowski space-time is   
\footnote{   
 We follow the convention   
in Ref.~\cite{Bergshoeff:1987cm}.   
}   
\begin{eqnarray}   
 I_P &=&   
 \displaystyle{   
  - \int d^3 \sigma \sqrt{-g} + I_{\rm WZ}   
   \, ,   
  \label{eq:Poly_action}   
 }   
\end{eqnarray}   
with the Wess-Zumino term $I_{\rm WZ}$ given by 
\begin{eqnarray}   
 I_{\rm WZ}   
 &=&   
 \displaystyle{   
  \int d^3 \sigma\,   
  \frac{i}{2} \epsilon^{ijk}   
   (\overline{\theta} \Gamma_{\mu\nu} \del_i \theta)   
   \left(   
    - \Pi_j^{\ \mu} \Pi_k^{\ \nu}   
    - i \Pi_j^{\ \mu}   
       (\overline{\theta} \Gamma^\nu \del_k \theta) 
   \right. 
 } \nonumber \\ 
 && \qquad \qquad \qquad \qquad \qquad 
 \displaystyle{ 
   \left. 
    + \frac{1}{3}   
       (\overline{\theta} \Gamma^\mu \del_j \theta)   
       (\overline{\theta} \Gamma^\nu \del_k \theta)   
   \right) \, .   
 }    
\end{eqnarray}   
 Here $g_{ij}$ is   
\begin{equation}   
 g_{ij} = \Pi_i^{\ \mu} \Pi_j^{\ \nu} \eta_{\mu\nu} \, ,   
\end{equation}   
with the invariant $1$-form $\Pi_i^{\ \mu}$ defined by   
\begin{equation}   
 \Pi_j^{\ \mu}   
 =   
 \del_j X^\mu - i \bar{\theta} \Gamma^\mu \del_j \theta \, .   
\end{equation}   
 $\theta$ is a Majorana fermion and $\bar{\theta} = \theta^T C$   
where $C$ is the charge conjugation matrix 
in $11$ dimension. 
 The membrane tension is set equal to 1 here.   
  
 In this paper, 
we would like to study the membrane with its one spatial direction, 
say, $\sigma^2 \approx \sigma^2 + 2\pi$, 
wrapped once along the eleventh coordinate 
$x^{\#} \approx x^{\#} + 2 \pi R$. 
 The membrane coordinate $X^{\#}$ satisfies   
\begin{equation}   
 X^{\#}(\sigma^0,\,\sigma^1,\,\sigma^2 + 2\pi)   
 =   
 X^{\#}(\sigma^0,\,\sigma^1,\,\sigma^2) + 2\pi R \, .   
\end{equation}   
 Such a coordinate can be rewritten as   
\begin{equation}   
 X^{\#}(\sigma)   
 = R \left( \sigma^2 - A_1 (\sigma) \right) \, ,   
  \label{eq:X_A}   
\end{equation}   
by using $A_1$ obeying the periodic boundary condition   
\begin{equation}   
 A_1(\sigma^0,\,\sigma^1,\,\sigma^2 + 2\pi)   
 =   
 A_1(\sigma^0,\,\sigma^1,\,\sigma^2) \, .   
\end{equation}   
 From the action, one can obtain the conjugate momenta $P_\mu$, $P_{A_1}$   
and $\chi_\alpha$ of $X^\mu$ 
($\mu = 0, \cdots, 9$), $A_1$ and $\theta^\alpha$ 
respectively.   
 It is also easy to identify 
the Hamiltonian constraint $\phi_0 \approx 0$   
and the momentum constraints $\phi_r \approx 0$ ($r = 1,\,2$)   
as well as the fermionic constraints $\xi_\alpha \approx 0$ 
corresponding to the kappa symmetry.   
  
 As usual, we decompose   
$X^\mu$, for instance, into the light-cone directions 
\begin{equation} 
 X^\pm = \frac{1}{\sqrt{2}} \left( X^9 \pm X^0 \right) \, , 
\end{equation} 
and the transverse directions $X^I$ ($I = 1, \cdots, 8$). 
 Now we employ the light-cone gauge fixing 
\begin{eqnarray}   
 && 
 P^+ = {\rm constant} \, , 
  \nonumber \\ 
 && 
 X^+ = \sigma^0 \, ,   
  \nonumber \\   
 &&   
 \Gamma^+ \theta = 0 \, .   
\end{eqnarray}   
 By using the representation for gamma matrices   
\begin{eqnarray}  
 &   
 \displaystyle{   
  \Gamma^0 = i \sigma_2 \otimes {\bf 1}_{16} = C \, ,   
  \quad     
  \Gamma^{\#} = \sigma_3 \otimes \gamma^9 \, ,   
 }& \nonumber \\   
 &   
 \displaystyle{   
  \Gamma^I = \sigma_3 \otimes \gamma^I \, , \quad   
  \Gamma^9 = \sigma_1 \otimes {\bf 1}_{16} \, ,   
 }&   
\end{eqnarray}    
where $\gamma^j$ ($j=1, \cdots, 9$) 
are the gamma matrices in $9$-dimensional Euclidean space,   
together with an appropriate rescaling, 
we get  
the action which looks like that of a two dimensional gauge theory 
as 
\begin{eqnarray}   
 I_0^{\rm lc}   
 &=&   
 \int dt d^2 \vec{\sigma}   
 \left[   
    \frac{R^3}{2} (F_{t1})^2   
  + \frac{1}{2} (D_t X^I)^2   
  - \frac{1}{2} (D_1 X^I)^2   
  - \frac{1}{4} \frac{1}{R^3} \left( \{ X^I,\,X^J \} \right)^2   
 \right.   
  \nonumber \\   
 && \qquad \qquad   
 \left.   
  - \frac{i}{2}\,\Psi^T D_t \Psi   
  - \frac{i}{2}\,\Psi^T \gamma^9 D_1 \Psi  
  + i \frac{1}{R^{3/2}}\, \frac{1}{2}\,   
    \Psi^T \gamma^I \{ X^I,\, \Psi \}   
 \right] \, .   
  \label{eq:lc_two_gauge}   
\end{eqnarray}   
 Here the time-like coordinate $t$ on the world volume is 
defined as 
\begin{equation} 
 t \equiv R \sigma^0 \, , 
  \label{eq:t_sigma_0}
\end{equation} 
and $\Psi$ denotes the fermionic variables left after the light-cone 
gauge fixing; 
\begin{equation} 
 \theta 
 = 
 \frac{1}{2^{\frac{3}{4}} \sqrt{P^+}} 
 \left( 
  \begin{array}{c} 
   \Psi \\ 
   0 
  \end{array} 
 \right)\, . 
\end{equation} 
 The bracket $\{, \}$ denotes   
\begin{equation}   
 \{ \Phi_1,\, \Phi_2 \} = \epsilon^{rs} \del_r \Phi_1 \del_s \Phi_2   
  \, ,   
\end{equation}   
with $\epsilon^{rs} = - \epsilon^{sr}$, $\epsilon^{12} = 1$,   
while the covariant derivative and the field strength are defined by   
\begin{eqnarray}   
 D_a \Phi &\equiv&   
  \del_a \Phi + \{ A_a,\, \Phi \} \quad (a = 0,\,1) \, ,   
  \nonumber \\   
 F_{01} &\equiv&   
  \del_0 A_1 - \del_1 A_0 + \{ A_0,\, A_1 \} \, .   
\end{eqnarray}   

 One important remark here is   
that the kinetic terms of 
$X^I$ do not involve the $\sigma^2$ derivatives.   
 In the case of the unwrapped membrane,   
the kinetic terms do not include the $\sigma^1$ and $\sigma^2$    
derivatives. 
 We need some regularization to make sense out of such 
a theory.  
 The matrix regularization   
replaces the dependence on the spatial coordinates    
by matrix degrees of freedom,   
and the regularized theory becomes a matrix quantum mechanics.   
 In our case, we expect that   
some two dimensional theory should regularize the wrapped membrane.   
  
\subsection{Sekino-Yoneya procedure}   
\label{subsec:sy_proc}   
 Sekino and Yoneya found a way   
to map the variables of the wrapped membrane action   
(\ref{eq:lc_two_gauge}) to the variables in matrix string theory.   
 Let us define $\sigma \equiv \sigma^1$ and $\rho \equiv \sigma^2$.   
 Now we will discretize the coordinate $\rho$, which corresponds to the   
wrapped direction, so that it takes only the following values:   
$\rho \in \{0, \frac{2\pi}{N}, \cdots, \frac{2\pi}{N} (N-1)\}$.   
 For each variable $\Phi (\sigma ,\rho )$ appearing in the membrane   
action, we Fourier decompose it as   
\begin{equation}   
 \Phi (\sigma,\,\rho)   
 = \sum_n e^{in \rho}\, \Phi_n(\sigma)\, .    
\end{equation}   
 Since $\rho$ is discretized, $n$ is restricted to   
$\left| n \right| \le \frac{N}{2}$.   
 From such Fourier modes, 
one can define an $N \times N$ matrix variable   
$\phi_{kl}(\theta)$ ($k,l=1,\cdots ,N$)   
following the boundary condition   
\begin{equation}   
 \phi_{kl}(\theta + 2\pi) = \phi_{k+1,\,l+1}(\theta) \, .   
  \label{eq:bc_matrix}   
\end{equation}   
 The correspondence between $\Phi_n(\sigma)$    
and    
$\phi_{kl}(\theta)$   
is given as   
\cite{Sekino:2001ai}    
\begin{equation}   
 \Phi_{k-l}(\sigma_{kl}(\theta)) = \phi_{kl}(\theta)\, ,    
  \label{eq:SY}   
\end{equation}   
where $\sigma_{kl}(\theta)$ is given by   
\footnote{   
 For finite $N$,   
the precise form of the mapping rule   
depends on   
whether $N$ is even or odd,   
$n$ is smaller, equal or larger than $N/2$ \cite{Sekino:2001ai}.   
 But any of its details are not necessary 
to taken into account here explicitly.   
}    
\begin{equation}   
 \sigma_{kl}(\theta)   
 \equiv \frac{(k-1)+(l-1)}{N}\,\pi + \frac{\theta}{N} \, .   
  \label{eq:SY_rule}   
\end{equation}   
 Therefore, in the matrix description, the $\sigma$-direction on the   
world volume of the membrane is divided   
into $N$ segments and each matrix element corresponds to each segment.   
 The Kaluza-Klein modes ($n \ne 0$) along the wrapped direction   
$\rho$ are packed   
into the off-diagonal elements of the matrix variables   
($k \ne l$).   
   
 To see that the light-cone membrane action (\ref{eq:lc_two_gauge})   
is indeed mapped to the matrix string action,   
it is necessary to rewrite   
the bracket $\{,\}$ and the spatial integral    
$\frac{2\pi}{N} \sum_\rho \int d\sigma$   
in terms of matrices.   
 It is straightforward to check the correspondence    
\begin{eqnarray}   
 &&   
 \displaystyle{   
  \frac{1}{N} \sum_\rho e^{-i(k-l)\rho} \{ \Phi_1,\,\Phi_2 \}   
  \mapsto   
  i \left( \frac{N}{2\pi} \right)   
  [\phi_1,\,\phi_2]_{kl} \, ,   
 }   
  \nonumber \\   
 &&   
 \displaystyle{   
  \frac{1}{N} \sum_\rho \int_0^{2\pi} \frac{d\sigma}{2\pi}   
  \mapsto   
  \int_0^{2\pi} \frac{d\theta}{2\pi}\, \frac{1}{N} {\rm tr}\, ,   
 }   
  \nonumber \\   
 &&   
 \del_\sigma \mapsto N \del_\theta \, ,   
  \nonumber \\   
 &&   
 \displaystyle{   
  A_\sigma \mapsto a_\theta \, ,   
 }   
  \label{eq:map_rule}   
\end{eqnarray}   
which is valid in the large $N$ limit.   
 By using those relations,   
the action (\ref{eq:lc_two_gauge}) is replaced by   
\begin{eqnarray}   
 I_{\rm WM} 
 &=& 
 \int dt \int_0^{2\pi} \frac{d\theta}{2\pi} 
 \frac{1}{N} {\rm tr} 
 \left[   
    \frac{R^3}{2} (f_{t\,\theta})^2   
  + \frac{1}{2} (D_t x^I)^2   
 \right.   
  \nonumber \\   
 && \qquad \qquad \qquad \quad   
 \left.   
  - \frac{N^2}{2} (D_\theta x^I)^2   
  - \frac{N^2}{(2\pi)^2 R^3}   
    \frac{1}{4} \left( i[x^I,\,x^J] \right)^2   
 \right.   
  \nonumber \\   
 && \qquad \qquad \qquad \quad   
 \left.   
  - \frac{i}{2}\,\psi^T \rho^0 D_t \psi   
  - N \frac{i}{2}\, \psi^T \gamma^9 D_\theta \psi   
 \right.   
  \nonumber \\   
 && \qquad \qquad \qquad \quad   
 \left.   
  - \frac{N}{2\pi R^{3/2}}\, \frac{1}{2}\,    
    \psi^T \gamma^I [x^I,\,\psi]   
 \right] \, ,   
  \nonumber \\   
  \label{eq:MS_action_0}   
\end{eqnarray}   
with   
\begin{equation}   
 f_{t\,\theta}   
 \equiv \drv{t} a_\theta - N \drv{\theta} a_t   
 + i \frac{N}{2\pi}\,[a_t,\,a_\theta] \, .   
\end{equation}    
 By using the rescaled time-like coordinate $\tau$ defined by   
\begin{equation}   
 \tau \equiv N\,t \, ,   
  \label{eq:t_tau}   
\end{equation}   
the above action is rewritten as   
\begin{eqnarray}   
 I_{\rm MS}   
 &=&   
 \int d\tau \int_0^{2\pi} \frac{d\theta}{2\pi}\,   
 {\rm tr}   
 \left[   
    \frac{R^3}{2} (f_{\tau\,\theta})^2   
  + \frac{1}{2} (D_\tau x^I)^2   
 \right.   
  \nonumber \\   
 && \qquad \qquad \qquad \quad   
 \left.   
  - \frac{1}{2} (D_\theta x^I)^2   
  - \frac{1}{(2\pi)^2 R^3}   
    \frac{1}{4} \left( i[x^I,\,x^J] \right)^2   
 \right.   
  \nonumber \\   
 && \qquad \qquad \qquad \quad   
  \left.   
  - \frac{i}{2}\,\psi^T D_\tau \psi   
  - \frac{i}{2}\,\psi^T \gamma^9 D_\theta \psi 
  \right.   
  \nonumber \\   
 && \qquad \qquad \qquad \quad   
 \left.   
  - \frac{1}{2\pi R^{3/2}}\,\frac{1}{2}\, 
     \psi^T \gamma^I [x^I,\,\psi]   
 \right] \, ,   
  \label{eq:MS_action}   
\end{eqnarray}   
where the covariant derivatives and $f_{\tau\,\theta}$ are 
respectively 
\begin{eqnarray}   
 &&   
 \displaystyle{   
  f_{\tau\,\theta} \equiv   
    \drv{\tau} a_\theta - \drv{\theta} a_\tau   
  + i \frac{1}{2\pi} \, [a_\tau,\,a_\theta ] \, ,   
 } \nonumber \\   
 &&   
 a_\tau \equiv a_t\, ,   
  \nonumber \\   
 &&   
 \displaystyle{   
  D_a \phi   
  \equiv   
  \drv{a} \phi + i\frac{1}{2\pi}\,[a_a,\,\phi]   
  \quad (a = \tau,\,\theta)\, .   
 }   
  \label{eq:MS_theory}   
\end{eqnarray}   
 The action (\ref{eq:MS_action}) coincides with the action   
of matrix string theory \cite{matrix_string}.   
 Hence, the quantum wrapped membrane will be defined   
by matrix string theory   
with the matrix variables obeying the twisted boundary condition   
(\ref{eq:bc_matrix}),   
instead of the periodic one   
used in the original matrix string theory   
\cite{matrix_string}.   

\section{Sekino-Yoneya procedure from $D$-branes}   
\label{sec:d1}   
 In this section, we would like to rederive the Sekino-Yoneya procedure   
using the M(atrix) theory techniques.   
 We start from matrix string theory.   
 Matrix string theory is basically the theory of $D0$-branes   
with one space-like direction compactified.   
 It was originally invented 
by following Fig.~\ref{fig:duality_chain}. 

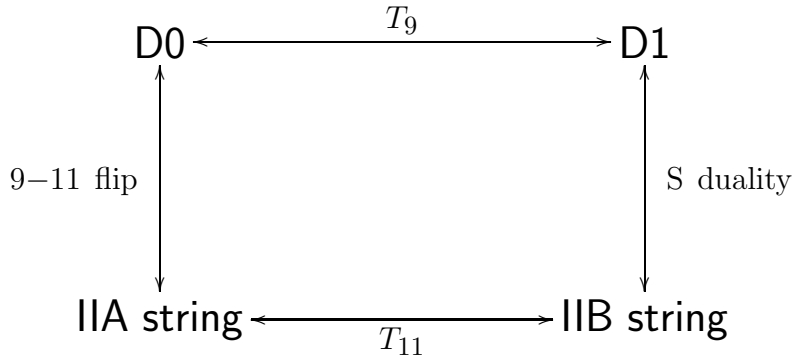
\begin{figure} 
 \caption{Connection between Type IIA strings 
         and $D0$-branes via duality transformations.   
         $T_\mu$ denotes the $T$-dual transformation along   
         the $x^\mu$-direction.} 
\begin{eqnarray} 
\Large 
\xymatrixrowsep{30mm} 
\xymatrixcolsep{40mm} 
\xymatrix{ 
  {\sf D0}  
  \ar@{<->}[r]^{T_9} 
  \ar@{<->}[d]_{{\rm 9-11\ flip\ }} & 
  {\sf D1} 
  \ar@{<->}[d]^{{\rm \ S\ duality}} \\ 
  {\sf IIA\ string} 
  \ar@{<->}[r]_{T_{11}} & 
  {\sf IIB\ string} 
} 
 \nonumber 
\end{eqnarray} 
 \label{fig:duality_chain} 
\end{figure} 

 Since Type IIA strings can be considered as wrapped membranes,   
the matrix regularization by Sekino and Yoneya may be derived 
by closely examining this figure.   
 We will show that this is indeed the case and 
the variables on the wrapped membrane are 
precisely expressed by the rules (\ref{eq:SY}),   
(\ref{eq:SY_rule}) in terms of the matrix string variables.   
  
 Let us start with a collection of $MN$ $D0$-branes.   
 The variables on the world volume corresponding to such branes   
are $MN\times MN$ hermitian matrices.   
 An $MN\times MN$ matrix $F$ can be expressed as   
\begin{equation}   
 F = \sum_{\vec{m}} f_{\vec{m}}\,J^{(MN)}_{\vec{m}} \, ,   
\end{equation}   
where 
$\{ J^{(MN)}_{\vec{m}} \}_{\vec{m}}$ is a basis   
of $MN\times MN$ matrices defined by Eq.~(\ref{eq:J})   
in Appendix,   
and $\vec{m} \equiv (m_1, m_2)$ with    
$-\frac{MN}{2} < m_1, m_2 \le \frac{MN}{2}$.   
 Considering $J^{(MN)}_{\vec{m}}$ to be a regularized version   
of $e^{im_1\sigma^1 +im_2\sigma^2 }$,   
we can get the matrix regularization.   
  
 We would like to express a membrane wrapped around a compactified   
direction $x^9$ using these $D0$-branes.   
 We will regard these $MN$ $D0$-branes as $M$ copies of $N$ $D0$-branes,   
consider these copies as the mirror images 
due to the torus compactification,   
and take the limit $M\rightarrow\infty$ eventually.   
 Accordingly, we express an $MN\times MN$ matrix $F$ in the form of   
an $M\times M$ matrix $F_{ab}$, $(a,b=1,\cdots ,M)$ with each   
matrix element $F_{ab}$ being an $N\times N$ matrix.   
 Thus the matrix variables $A_0(\tau)$, $x^I(\tau)$ ($I=1, \cdots, 8$),   
$x^9(\tau)$ (which becomes the extra dimension $x^{\#}$   
after the $9 \leftrightarrow \#$ flip),   
$\psi(\tau)$ on the world volume satisfies   
the ${\bf Z}$-orbifolding condition \cite{Taylor:1996ik}   
\begin{eqnarray}   
 &   
 \displaystyle{   
  \phi_{ab}(\tau) = \phi_{a-1,\,b-1}(\tau)  \, ,   
 }& \nonumber \\   
 &   
 \displaystyle{   
  x^9_{ab}(\tau) = x^9_{a-1,\,b-1}(\tau) + 2\pi R \delta_{ab} \, ,  
 }&   
 \label{eq:BC_D0}   
\end{eqnarray}   
where we express the variables other than $x^9$ as   
$\phi$ collectively.    
 The configuration of a membrane wrapped around the   
compactified direction $x^9$ can be expressed as   
$x^9(\tau) = M R Q_{(MN)}$,   
where $Q_{(MN)}$ is an $MN \times MN$ matrix defined by   
\begin{equation}   
 Q_{(MN)}   
 =   
    \left(   
     \begin{array}{ccccccc}   
      0 & & & & & & \\   
      & \frac{2\pi}{MN} & & & & & \\   
      & & \cdot & & & & \\   
      & & & \cdot & & & \\   
      & & & & \cdot & & \\   
      & & & & & \cdot & \\   
      & & & & & & \frac{2\pi}{MN} (MN-1)   
     \end{array}   
    \right) \, .   
  \label{eq:QMN}  
\end{equation}   
 This obviously satisfies the boundary condition   
in Eq.~(\ref{eq:BC_D0}).   
 Moreover, since $e^{iQ_{(MN)}}=V_{(MN)}\sim e^{i\sigma^2}$,   
this configuration can be considered as the matrix version   
of $x^9(\tau) = R \rho $ with $\rho =M\sigma^2$.   
 Thus we take the variable $x^9(\tau)$ to   
be the sum of the background and the fluctuations   
around it:  
\begin{eqnarray}   
 &   
 \displaystyle{   
  x^9(\tau) = M R Q_{(MN)} + \hat{x}^9(\tau) \, ,   
 }& \nonumber \\   
 &   
 \displaystyle{   
  \hat{x}^9_{ab}(\tau) = \hat{x}^9_{a-1,\,b-1}(\tau) \, .   
 }&   
 \label{eq:QMN+xhat}  
\end{eqnarray}   

 Now a matrix $F = \sum_{\vec{m}} f_{\vec{m}}\,J^{(MN)}_{\vec{m}}$   
is interpreted as a function   
$f(\sigma ,\rho )$ depending on two coordinates   
$\sigma ,\rho $ defined as   
\begin{equation}   
 f(\sigma ,\rho )   
 = \sum_{\vec{m}} f_{\vec{m}}\,e^{im_1\sigma +im_2^\prime \rho } \, ,  
\label{eq:fsigmarho}  
\end{equation}   
with $m_2^\prime =m_2/M$ and $\sigma =\sigma^1$.   
 As we will see, the condition (\ref{eq:QMN+xhat}) requires   
that $m_2$ is divisible by $M$.   
 Therefore, in the expansion (\ref{eq:fsigmarho}),   
$\rho$ can be considered to be discretized in unit of $\frac{2\pi}{N}$.   
 Under this rule,   
the commutators of matrix variables are mapped to   
the brackets $\{,\}$ with a normalization factor.   
  
 In order to get the matrix string variables, we need to T-dualize   
the above matrix variables.   
 Following the familiar procedure   
\cite{Taylor:1996ik},   
we introduce $N \times N$ matrices $\widetilde{\Phi}(\tau,\,\tilde{x} )$,   
$\widetilde{A}_{\tilde{x}}(\tau,\,\tilde{x} )$ as   
\begin{eqnarray}   
 &   
 \displaystyle{   
  \widetilde{\Phi}(\tau,\,\tilde{x} )   
  \equiv   
  \sum_a e^{ia\tilde{x} /\widetilde{R}} \phi_{0 a}(\tau) \, ,   
 }& \nonumber \\   
 &   
 \displaystyle{   
  \widetilde{A}_{\tilde{x}}(\tau,\,\tilde{x})   
  \equiv   
  \sum_a   
    e^{ia\tilde{x} /\widetilde{R}}   
    \left( \frac{1}{2\pi\alpha^\prime} x^9_{0 a}(\tau) \right)   
   \, .   
 }&   
  \label{eq;T_dual_tr}   
\end{eqnarray}   
 Here $\tilde{x},~(\tilde{x}\approx \tilde{x}+2\pi\widetilde{R})$   
is the coordinate on the dual torus and   
$\widetilde{R}\equiv \alpha^\prime /R$.  
 In this representation,   
$x^9(\tau)$ is replaced by a differential operator   
$2\pi \alpha^\prime (i \del_{\tilde{x}}   
+ \widetilde{A}_{\tilde{x}}(\tau,\,\tilde{x})) $.   
 Usually $\widetilde{A}_{\tilde{x}}$, $\widetilde{\Phi}$   
thus obtained yield the variables in matrix string theory.   
 In our case,   
from Eq.~(\ref{eq:QMN+xhat}), we get   
$x^9_{00}(\tau) = R Q_{(N)} + \hat{x}^9_{00}(\tau)$,   
where $Q_{(N)}$ is given by Eq.~(\ref{eq:QMN}) with   
$MN$ replaced by $N$.   
 Thus $\widetilde{A}_{\tilde{x}}$ defined above contains   
a background piece proportional to $Q_{(N)}$.   
 This background can be eliminated by a global gauge   
transformation, which changes the boundary conditions   
of the variables. Namely we define the new variables   
as follows:   
\begin{eqnarray}   
 &   
 \displaystyle{   
  A_{\tilde{x}}   
  =   
  \exp   
  \left( -i \frac{\tilde{x}}{2\pi\widetilde{R}} Q_{(N)} \right)   
  \left( i \del_{\tilde{x}} + \widetilde{A}_{\tilde{x}} \right)   
  \exp   
  \left( i \frac{\tilde{x}}{2\pi\widetilde{R}} Q_{(N)} \right) \, ,   
 }& \nonumber \\   
 &   
 \displaystyle{   
  \Phi   
  =   
  \exp   
  \left( -i \frac{\tilde{x}}{2\pi\widetilde{R}} Q_{(N)} \right)   
  \widetilde{\Phi}   
  \exp   
  \left( i \frac{\tilde{x}}{2\pi\widetilde{R}} Q_{(N)} \right) \, .   
 }&   
  \label{eq:twisted_gt}   
\end{eqnarray}   
 In order to compare the notation used in the previous section,   
we define    
\begin{equation}   
 \tilde{x} = \widetilde{R} \theta \, , \quad   
 A_{\tilde{x}}(\tilde{x}) = \frac{1}{\widetilde{R}} A_\theta(\theta) \, ,   
\end{equation}   
and consider $A_\theta$, $\Phi$ to be functions of   
$\theta$.   
 $A_\theta$, $\Phi$ obey the twisted boundary condition:   
\begin{eqnarray}   
 &   
 \displaystyle{   
  A_\theta(\theta + 2\pi)   
  = (V_{(N)})^{-1} A_\theta(\theta)\, V_{(N)} \, ,   
 }& \nonumber \\   
 &   
 \displaystyle{   
  \Phi(\theta + 2\pi)   
  = (V_{(N)})^{-1} \Phi(\theta)\, V_{(N)} \, ,  
 }&   
  \label{eq:twisted_bc}   
\end{eqnarray}   
where $V_{(N)}=\exp (iQ_{(N)})$.   
 By a unitary transformation   
$A_\theta \mapsto (S_{(N)})^\dagger A_\theta S_{(N)}$ and    
$\Phi \mapsto (S_{(N)})^\dagger \Phi S_{(N)}$,   
with    
\begin{equation}   
 S_{(N)}   
 =   
 \frac{1}{\sqrt{N}}   
 \left(   
  \begin{array}{ccccccc}   
   1 & 1 & 1 & \cdot & \cdot & \cdot & 1 \\   
   1 & \omega_{(N)} & \omega_{(N)}^2   
   & \cdot & \cdot & \cdot & \omega_{(N)}^{N-1} \\    
   \cdot & \cdot & \cdot & \cdot & \cdot & \cdot & \cdot \\   
   \cdot & \cdot & \cdot & \cdot & \cdot & \cdot & \cdot \\   
   \cdot & \cdot & \cdot & \cdot & \cdot & \cdot & \cdot \\   
   1 & \omega_{(N)}^{N-1}   
   & \omega_{(N)}^{2(N-1)} & \cdot & \cdot & \cdot &   
     \omega_{(N)}^{(N-1)^2} \\    
  \end{array}   
 \right) \, ,   
  \label{eq:S_N}   
\end{equation}   
satisfying   
\begin{equation}   
 \quad   
 \left\{   
  \begin{array}{l}   
   (S_{(N)})^\dagger V_{(N)} S_{(N)} = (U_{(N)})^{-1} \\   
   (S_{(N)})^\dagger U_{(N)} S_{(N)} = V_{(N)}   
  \end{array}   
 \right. \, ,    
\end{equation}   
we can turn the boundary condition into   
\begin{eqnarray}   
 &   
 \displaystyle{   
  A_\theta(\theta + 2\pi)   
  = U_{(N)} A_\theta(\theta)\, (U_{(N)})^{-1} \, ,   
 }& \nonumber \\   
 &   
 \displaystyle{   
  \Phi(\theta + 2\pi)   
  = U_{(N)} \Phi(\theta)\, (U_{(N)})^{-1} \, ,  
 }&   
\end{eqnarray}   
which coincide with the one (\ref{eq:bc_matrix})   
satisfied by the matrix string variables   
appeared in the previous section.   
  
 Now what we should do is to show how the variables obtained   
via the above procedure is related to the world volume variables   
on the membranes, simply following what we have done.   
 Let $F_{rs},~(r,s=1,\cdots ,MN)$ denotes the matrix element   
of an $MN\times MN$ matrix $F$.   
 From Eq.~(\ref{eq:BC_D0}),   
we need matrices $F$ satisfying the condition   
\begin{equation}   
 F_{r+N,\,s+N} = F_{rs} \, ,  
  \label{eq:cond_F}   
\end{equation}   
to represent the configurations of $D0$-branes.   
 Since the matrix elements of $J^{(MN)}_{\vec{m}}$ are   
\begin{equation}   
 \left( J^{(MN)}_{\vec{m}} \right)_{rs}   
 =   
 \left( \omega_{MN} \right)^{m_2(-\frac{m_1}{2} + s - 1 )}   
 \delta^{(MN)}_{r-s,\,-m_1} \, ,   
\end{equation}   
where   
\begin{equation}   
 \delta^{(MN)}_{rs}   
 =   
 \left\{   
  \begin{array}{ll}   
   1 & r\equiv s \ \mbox{mod}\ MN \\   
   0 & \mbox{otherwise}   
  \end{array}   
 \right. \, ,   
\end{equation}   
the condition (\ref{eq:cond_F})   
for $F = J^{(MN)}_{\vec{m}}$ is satisfied if and only if   
$m_2$ is divisible by $M$, i.e.,   
$m_2 = m_2^\prime M$ with $-\frac{N}{2} < m_2^\prime \le \frac{N}{2}$.   
 Hence, $F$ obeying the condition (\ref{eq:cond_F})   
is in general written by   
\begin{equation}   
 F = \sum_{m_1,\,m_2^\prime} f_{(m_1,\,m_2^\prime M)}\,J^{(MN)}_{(m_1,\,m_2^\prime M)} \, ,   
\label{eq:Fper}  
\end{equation}   
where the sum runs over $-\frac{N}{2} < m_2^\prime \le \frac{N}{2}$,   
and $-\frac{MN}{2} < m_1 \le \frac{MN}{2}$.   
  
 After the T-dualization (\ref{eq;T_dual_tr}), we should take   
$N\times N$ matrices   
$\{ J^{(N)}_{\vec{m}}(\theta) \}_{\vec{m}}$ whose   
$(k,\,l)$-element ($k, l = 1, \cdots, N$) is    
\begin{equation}   
 \left( J^{(N)}_{\vec{m}}(\theta) \right)_{kl}   
 \equiv   
 \sum_a e^{i a \theta}   
  \left( J^{(MN)}_{\vec{m}} \right)_{k,\, l + aN } \, ,   
 \label{eq:J_theta}   
\end{equation}   
as the basis for the matrix string variables   
satisfying the periodic boundary conditions.   
 Accordingly the $MN\times MN$ matrix $F$ in   
Eq.~(\ref{eq:Fper}) is mapped to   
\begin{eqnarray}   
 &&   
 \displaystyle{   
  F = \sum_{m_1,\,m_2^\prime}   
      f_{(m_1,\,m_2^\prime M)}\,J^{(MN)}_{(m_1,\,m_2^\prime M)}   
 } \nonumber \\   
 && \qquad \mapsto   
 \displaystyle{   
  \tilde{f}(\theta)   
  = \sum_{m_1,\,m_2^\prime}   
    f_{(m_1,\,m_2^\prime M)}\,  
    J^{(N)}_{(m_1,\,m_2^\prime M)}(\theta) \, ,  
 }   
  \label{eq:tilde_f_theta}   
\end{eqnarray}   
whose matrix elements are calculated as   
\begin{equation}   
 \tilde{f}_{kl}(\theta)    
 =  
  \sum_{m_1,\,m_2^\prime}    
  f_{(m_1,\,m_2^\prime M)}  
  \sum_a   
  \delta_{k-l-aN,\,-m_1}  
  \exp\left(   
      \frac{2\pi i}{N} m_2^\prime \left(\frac{m_1}{2} + k - 1\right)   
      +ia\theta   
      \right) .  
  \label{eq:tilde_f_theta_kl}   
\end{equation}   
 Here the range of summation over $a$ is chosen so that   
$-\frac{MN}{2} < m_1=l-k+aN \le \frac{MN}{2}$.    
  
 To map $\tilde{f}(\theta)$ to the field $f(\theta)$   
obeying the twisted boundary condition in Eq.~(\ref{eq:twisted_bc}),   
we apply the transformation   
in Eq.~(\ref{eq:twisted_gt}) to $\tilde{f}(\theta)$ and we get  
\begin{equation}   
 f(\theta)   
 =    
   e^{-i \frac{\theta}{2\pi} Q_{(N)}}   
   \tilde{f}(\theta)   
   e^{ i \frac{\theta}{2\pi} Q_{(N)}}    
\end{equation}   
 Substituting Eq.~(\ref{eq:tilde_f_theta_kl}) into this,   
we can get the matrix element of $f(\theta )$.   
 Since $-\frac{MN}{2} < m_1 \leq \frac{MN}{2}$ can be written as   
$m_1=m_1^\prime +a^\prime N$   
with $-\frac{N}{2} < m_1^\prime \leq \frac{N}{2}$ and   
$-\frac{M}{2} < a^\prime \leq \frac{M}{2}$,   
we can decompose the summation over $m_1$   
into those of $m_1^\prime$ and $a^\prime$ and obtain   
\begin{equation}   
  f_{kl}(\theta)   
 =  
  \sum_{m_1^\prime ,\,m_2^\prime}  
  \left(  
  \sum_{a^\prime}   
  f_{(m_1^\prime +a^\prime N,\, m_2^\prime M)}  
  e^{ia^\prime (\theta + \pi m_2^\prime )}  
  \right)  
  \left(  
  J_{\vec{m}^\prime }^{(N)}  
  \right)_{kl}  
  e^{\frac{i}{N}m_1^\prime \theta} .  
 \label{eq:f_theta_1}  
\end{equation}  
 Performing the unitary transformation $S_{(N)}$ in Eq.~(\ref{eq:S_N}),   
we get    
\begin{eqnarray}   
 f_{kl}(\theta)   
 &=&  
  \sum_{m_1^\prime ,\,m_2^\prime}  
  \sum_{a^\prime}   
  f_{(m_1^\prime + a^\prime N,\, m_2^\prime M)}  
  \delta^{(N)}_{k-l,m_2^\prime}  
  \exp   
  \left(   
   \frac{i}{N}(m_1^\prime +a^\prime N)   
              \left(\pi (-m_2^\prime + 2k-2) +\theta\right)   
  \right)    
  \nonumber  
  \\  
  &=&  
  \sum_{m_1,\,m_2^\prime}  
  f_{(m_1,\,m_2^\prime M)}  
  \delta^{(N)}_{k-l,m_2^\prime}  
  \exp   
  \left(   
   \frac{i}{N}m_1\left(\pi (-m_2^\prime + 2k-2) +\theta\right)   
  \right) \, .  
\end{eqnarray}  
 Since the summand depends only on the combination   
$m_1=m_1^\prime +a^\prime N$, the summation over $m_1^\prime$ and   
$a^\prime$ on the first line amounts to a summation over $m_1$.   
 Now let us compare this last expression with the function   
(\ref{eq:fsigmarho}) corresponding to the matrix variable   
in the limit $M\rightarrow\infty$.   
 Then we can read off $f_{kl}(\theta)$ as    
\begin{equation}   
 f_{kl}(\theta)   
 =   
 \frac{1}{N}   
 \sum_\rho   
 \sum_{m_2^\prime} e^{-i m_2^\prime \rho}\,    
 \delta^{(N)}_{k-l,\,m_2^\prime}   
 f\left(   
   \sigma = \frac{-m_2^\prime + 2 k - 2}{N} \pi + \frac{\theta}{N} \, ,   
   \rho   
  \right) \, .   
  \label{eq:rule}   
\end{equation}   
 In order to obtain Sekino-Yoneya mapping rule (\ref{eq:SY_rule}),    
we assume that $N$ is even and restrict $m_2$ to be even.   
 Then we get  
\begin{equation}   
 f_{kl}(\theta )  
 =  
 \frac{1}{N}\sum_{\rho}e^{-i(k-l)\rho}  
 f\left(   
   \sigma =\frac{k+l-2}{N}\pi +\frac{\theta}{N}\,,\,\rho   
   \right) \, ,   
 \end{equation}  
which coincides with the rule given by Sekino and Yoneya.   
 We note that   
the modulo $\pi$ periodicity along $\sigma$ is also derived   
to perform the sum over $m_2^\prime$ consistent   
with modulo $N$ periodicity.   
 Restricting $m_1$ to be even does not matter as a regularization,   
but it may cause some trouble because we should discard   
some of the degrees of freedom on the world volume of $D0$-branes.   
 One way to get the rule similar to Sekino-Yoneya  mapping rule   
without discarding the odd modes   
is to use a different basis of matrices:   
\begin{equation}   
 J^{\prime\,(N)}_{\vec{m}}   
  \equiv \left( \omega_{(N)} \right)^{-m_1 m_2}   
         \left( U_{(N)} \right)^{2m_1}   
         \left( V_{(N)} \right)^{m_2}   
  \, .  
   \label{eq:Jprime}   
\end{equation}   
 Assuming $N$ to be odd, $J^{\prime\,(N)}_{\vec{m}}$ can be used   
as a basis of $N\times N$ matrices.   
 Using $J^{\prime\,(N)}$   
in the above procedure assuming $N$ and $M$ to be odd,   
we get a mapping rule   
\begin{equation}  
 f_{kl}(\theta)  
 =  
 \frac{1}{N}\sum_{\rho}e^{-i(k-l)\rho}  
 f\left(   
   \sigma = 2\left(\frac{k+l-2}{N}\pi +\frac{\theta}{N}\right)\,,\,\rho   
  \right)  
\end{equation}  
 This relation gives a one-to-one correspondence at finite $M$,   
but after $M \rightarrow \infty$ to get matrix string theory,   
matrices becomes a double-cover of the membrane variables.   
   
\section{Super-Poincar\'e symmetry in the short distance regime}   
\label{sec:Lorentz}   
  
\subsection{remark on dynamics of regularized membrane}   
\label{sec:remark}   
  
 As we have established in the previous section, matrix string   
theory describes wrapped $M2$-branes in M theory in the DLCQ limit.   
 Here let us notice that   
the time coordinate $\tau$ used in the matrix string action   
(\ref{eq:MS_theory}) is not precisely $\sigma^0$ used 
in the membrane action.   
 From Eqs.~(\ref{eq:t_sigma_0}) and (\ref{eq:t_tau}),   
$\tau \sim RN \sigma^0$. 
 Thus, 
the natural time scale $\sigma^0\sim {\cal O}(1)$ 
of the membrane corresponds to $\tau \sim {\cal O}(NR)$.   
 Since we are interested in the limit $N\rightarrow\infty$,   
this fact implies that the very infrared dynamics of matrix string   
theory describes the world volume theory of wrapped membranes.   
 Matrix string theory is basically a two dimensional gauge theory,   
and it becomes strongly coupled in the infrared.   
 Hence, in order to describe wrapped membranes, we need to   
study the strongly coupled gauge theory.   
 Let us see this more explicitly in the following.   
  
 As a regularization of the world volume theory of membranes,   
all the variables are replaced by matrices in matrix string theory.   
 Products of matrices become products of corresponding variables   
in the limit $N\rightarrow \infty$ 
and commutators are proportional to brackets $\{,\}$.   
 For a finite $N$, the terms appearing in the matrix string action   
can be rewritten in terms of the variables appearing in the   
membrane action using the following formula \cite{Sekino:2001ai}:   
\begin{eqnarray}   
 &&   
 \int_0^{2\pi} \frac{d\theta}{2\pi}\,\frac{1}{N} {\rm tr}   
 \left( 
  \phi_{(1)}(\theta) \phi_{(2)}(\theta)  
  \cdots \phi_{(p)}(\theta) 
 \right) 
  \nonumber \\  
 && \quad   
 = \frac{1}{N} \sum_\rho \int_0^{2\pi} \frac{d\sigma}{2\pi}\, 
   \exp\left( 
        -i \frac{\pi}{N}   
        \sum_{1 \le j < k \le p}   
        \left(   
           \drv{\sigma_{(j)}} \drv{\rho_{(k)}}   
         - \drv{\rho_{(j)}} \drv{\sigma_{(k)}}   
        \right)   
       \right)   
  \nonumber \\   
 && \qquad \qquad \qquad   
 \times   
 \left.   
  \Phi_{(1)}(\sigma_{(1)},\,\rho_{(1)})   
   \Phi_{(2)}(\sigma_{(2)},\,\rho_{(2)})   
   \cdots \Phi_{(p)}(\sigma_{(p)},\,\rho_{(p)})   
 \right|_{\sigma_{(j)} = \sigma,\,\rho_{(j)} = \rho}  \, ,   
  \nonumber \\  
 && \quad  
 \equiv  
  \frac{1}{N} \sum_\rho \int_0^{2\pi} \frac{d\sigma}{2\pi}\,    
  \left( 
   \Phi_{(1)}\star \Phi_{(2)}\star \cdots \star \Phi_{(p)} 
  \right)(\sigma ,\,\rho),
\end{eqnarray}   
where $\drv{\rho}$ is a difference operator.   
 Using the noncommutative $\star$-product defined above,   
the action (\ref{eq:MS_action_0}) can be written in terms of   
the variables on the world volume of membranes as  
\begin{eqnarray}   
 I_{\rm WM}   
 &=&   
 \int dt \int_0^{2\pi} \frac{d\sigma}{2\pi} \frac{1}{N}\sum_\rho   
 \left[   
    \frac{R^3}{2} (F_{t\,\sigma})^2   
  + \frac{1}{2} (D_t X^I)^2   
 \right.   
  \nonumber \\   
 && \qquad \qquad \qquad \qquad   
 \left.   
  - \frac{1}{2} (D_\sigma X^I)^2   
  - \frac{N^2}{(2\pi)^2 R^3}\,  
    \frac{1}{4} \left( i[X^I,\,X^J] \right)^2   
 \right.   
  \nonumber \\   
 && \qquad \qquad \qquad \qquad 
  - \frac{i}{2}\, \Psi^T D_t \Psi   
  - \frac{i}{2}\, \Psi^T \gamma^9 D_\sigma \Psi   
  \nonumber \\   
 && \qquad \qquad \qquad \qquad   
 \left.   
  - \frac{N}{2\pi R^{3/2}}\, \frac{1}{2} 
    \Psi^T \gamma^I [X^I,\, \Psi]  
 \right]_\star \, .   
  \nonumber \\   
 F_{t\,\sigma}   
 &\equiv& \drv{t} A_\sigma - \drv{\sigma} A_t   
        + i \frac{N}{2\pi} [A_t,\,A_\sigma]_\star \, ,   
  \nonumber \\   
 D_a \Phi   
 &\equiv& \drv{a} \Phi + i \frac{N}{2\pi} [A_a,\,\Phi]_\star    
 \quad (a = t,\,\sigma) \, .   
  \label{eq:wm_star_action}   
\end{eqnarray}   
 After a rescaling $A_a \rightarrow A_a / R^{3/2}$   
to make the kinetic term for the gauge field canonical,   
the coupling constant of the theory regularizing the wrapped membrane   
is naively identified as   
\begin{equation}    
 \frac{N}{2\pi\,R^{3/2}} \, .    
\end{equation}   
 However, one should notice a subtlety coming from the fact that   
the kinetic terms in this action lack derivatives in the $\rho$-direction.   
 This implies that the free propagator   
of $X^\mu$ (and $A_t$, $A_\sigma$)   
depends on the discretized $\rho$-coordinate through the   
Kronecker-delta $\delta_{mn}$.   
 In the perturbative expansion using such propagators,   
one can easily see that the situation is similar to the   
large $N$ theory and the coupling constant becomes effectively   
the t'Hooft coupling  
\begin{equation}   
 g_{\rm WM}  
 = \frac{1}{2\pi} \left( \frac{N}{R} \right)^{3/2} \, .   
\end{equation}   

 On the other hand,   
the t'Hooft coupling constant   
in the matrix string action (\ref{eq:MS_action}) is   
identified as   
\begin{eqnarray}   
 g_{\rm MS}  
 &=& \frac{N^{1/2}}{2\pi R^{3/2}}   
  \nonumber \\   
 &=& \frac{g_{\rm WM}}{N} \, .    
\end{eqnarray}   
 The action in Eq.~(\ref{eq:wm_star_action}) was obtained by   
rewriting the regularized action (\ref{eq:MS_action}) using   
the $\star$-product.   
 The coupling constants of these theories have the dimension   
of the inverse of length and the relation between   
coupling constants reflects the fact   
that the coordinates $\tau ,\,\theta$ used in the matrix string action   
is obtained by dividing   
$\sigma^j$ used in the membrane action by $N$ essentially.    
 For the world sheet length scale $l$ of our interest,   
the dimensionless effective coupling constant $g_{\rm eff}$ is   
\begin{eqnarray}   
 g_{\rm eff} &=& l g_{\rm MS}   
  \nonumber \\   
 &=& \frac{l}{2\pi} \frac{N^{1/2}}{R^{3/2}} \, .   
\end{eqnarray}   
 Thus,   
as long as we are concerned with the length scale,   
\begin{equation}   
 l \le {\cal O}\left( 2\pi \frac{R^{3/2}}{N^{1/2}} \right) \, ,   
\end{equation}   
we can carry out perturbative expansion in $g_{\rm eff}$   
and observe the ultraviolet aspects   
of matrix string theory.   
  
 In the rest of this paper, we would like to examine if the   
super-Poincar\'e algebra closes for matrix string theory by studying this   
theory perturbatively.   
 Actually we only consider the lowest order   
in the perturbation theory.   
 Since this theory is supposed to describe M theory,   
we should require this as a consistency condition of the theory.   
 We cannot get any information 
on the super-Poincar\'e invariance 
of the gauge fixed world volume theory of 
the wrapped membranes from such analysis 
( See Ref.~\cite{B_membrane,S_membrane,LI_matrix}   
for the existence of critical dimension of the membrane   
and the related topics.), 
because we should look 
at the strongly coupled region of the theory to do so.    
 However, 
the correspondence between the membrane action and   
the matrix string action is useful 
when we try to find 
the super-Poincar\'e generators for matrix string theory, 
because, 
while matrix string theory itself does not give 
any intrinsic definition of such generators 
involving light-cone directions, 
the wrapped membrane action 
is classically super-Poincar\'e invariant.

\subsection{Matrix string in modified light-cone gauge}   
\label{sec:another}   
  
 Now let us study the matrix string action (\ref{eq:MS_action})   
at the lowest order in the perturbation theory. 
 Namely we put $g_{\rm MS}=0$. 
 In the wrapped membrane action (\ref{eq:wm_star_action}) 
equivalent to this action, 
putting $g_{\rm MS}=0$ corresponds to putting 
$g_{\rm WM}=0$. 
 Then there are no terms involving derivatives 
in the $\rho$ direction 
and we get $N$ independent strings 
lying on the world volume of the membrane.   
 Thus, naively, 
we can expect that the super-Poincar\'e generators are given 
as a sum of those for these strings.    
 However actually these strings are not completely independent in   
the light-cone gauge fixing described in Sec.~\ref{sec:membrane_matrix}.   
 Indeed $P^+$ is taken to be constant which means that each string   
carries the same amount of $p^+$. 
 Therefore the light-cone zero modes $p^+,\,x^-$ should be the   
same for each string, which makes it difficult to construct 
some of the super-Poincar\'e generators. 
 From the point of view of matrix string theory, 
this is consistent because there is only one $p^+$ 
and no notion of $p^+$ for each string. 
 However if one looks at the theory more closely, 
one can see that there exists degrees of freedom 
from which we can construct 
$p^+$ and $x^-$ for each string. 

 One may also worry about the pair $x^+,\,p^-$. 
 In the light-cone gauge, we should take the gauge so that $x^+$ 
is common to all the strings. 
 However this does not cause any problems in constructing the 
super-Poincar\'e generators, contrary to the $p^+,\,x^-$ case.  
  
 In this subsection, we would like to consider   
the {\it modified light-cone gauge},   
in which strings have independent $p^+,\,x^-$.   
 We will show that the variables in such a gauge can be   
written in terms of those in the light-cone gauge.   
 Thus the super-Poincar\'e generators in this new gauge can be expressed    
by the light-cone variables 
and we can calculate the commutators of these operators 
by using such expressions.   
  
 The action (\ref{eq:MS_action}) possesses the gauge symmetry   
\begin{equation}   
 \delta_\epsilon A_a   
 =  
 D_a\epsilon   
  \, ,    
\end{equation}  
which can be considered as a regularized version of the   
area preserving diffeomorphism.   
 The most convenient gauge choice would be   
$A_\sigma =0$, 
but since the $\sigma$-direction is compact,   
we cannot fix the modes in $A_\sigma$ 
which are constant with respect to $\sigma$.  
 Therefore the best we could do is to take $\partial_\sigma A_\sigma =0$  
so that we have degrees of freedom corresponding to $A_\sigma (\rho )$.   
 Actually we can reconstruct the variables corresponding to   
$p^+$ for each string from these degrees of freedom.   
  
 Similarly to matrix string theory, after taking the gauge   
$P^+=0$, we cannot fix $A_1=0$ using the area preserving diffeomorphism   
and the best we can do is to take $\partial_1A_1=0$.   
 In order to see which quantities should be identified as the   
$p^+$ for the strings, let us replace the gauge condition   
$P^+=0$ by $\partial_1P^+=0$.   
  This implies that $P^+ = P^+(\sigma^2)$ remains 
as a canonical variable 
and the strings put at the sites $\sigma^2$    
have independent longitudinal momenta $P^+(\sigma^2)$.   
 In this gauge, contrary to the light-cone gauge,   
we can take   
\begin{equation}   
 A_1(\vec{\sigma}) = 0 \, ,   
  \label{eq:A_1=0}   
\end{equation}   
using the residual diffeomorphism.   
  
 Thus, at least in the continuum membrane theory,   
we can take the modified light-cone gauge   
in which the strings put at the sites   
along the $\rho$ direction have independent $p^+$'s.   
 Since this is just another gauge fixing, 
we can express the variables in such a gauge 
in terms of the variables in the light-cone gauge. 
 Once such a relation is established, 
we will be able to find 
the super-Poincar\'e generators in the light-cone gauge 
from the ones constructed as a sum of those for independent strings 
in the modified light-cone gauge. 
  
 Classically it is straightforward to rewrite the variables   
in the modified light-cone gauge in terms of the variables in the   
light-cone gauge.   
 In order to distinguish the variables in two gauges,   
let $\phi_{\rm ml}(\sigma^1,\,\sigma^2)$ denote the   
variables in the modified light-cone gauge and   
$\phi_{\rm lc}(\sigma ,\,\rho )$ denote those in the light-cone gauge.   
 Here $\sigma^1=\sigma$ while $\sigma^2$ and $\rho$ are   
related through the equation   
\begin{equation}   
 \sigma^2 = \rho - A_{1\,{\rm lc}}(\rho) \, .   
  \label{eq:A_1}   
\end{equation}   
 $p^+_{\rm ml}(\sigma^2)$ can be obtained   
as   
\begin{equation}   
 p^+_{\rm ml}(\sigma^2)   
 =   
 \frac{P^+_{\rm lc}}   
 {1 - \del_\rho A_{1\,{\rm lc}}(\rho)}   
  \, .   
  \label{eq:P^+_rel}   
\end{equation}   
 $x^-_{\rm ml}(\sigma^2)$ which is the canonical conjugate   
of $p^+_{\rm ml}(\sigma^2)$ can be given as   
\begin{equation}   
 x^-_{\rm ml}(\sigma^2)   
 =   
 \int_0^{2\pi} \frac{d \sigma}{2\pi}   
  X^-_{\rm lc}(\sigma,\,\rho(\sigma^2)) \, ,   
   \label{eq:X^-_0}   
\end{equation}   
where $X^-_{\rm lc}(\sigma,\,\rho)$   
is given by solving the momentum constraints in the light-cone gauge.   
Other variables can be given as   
\begin{eqnarray}   
 &&   
 X^I_{\rm ml}(\sigma^1,\,\sigma^2) = X^I_{\rm lc}(\sigma^1,\,\rho) \, ,   
  \nonumber \\   
 &&   
 \theta_{\rm ml}(\sigma^1,\,\sigma^2) = \theta_{\rm lc}(\sigma^1,\,\rho) \, ,   
  \nonumber \\   
 &&   
 P^I_{\rm ml}(\sigma^1,\,\sigma^2)   
 = P^I_{\rm lc}(\sigma^1,\,\rho)   
   \frac{1}{1 - \del_\rho A_{1\,{\rm lc}}(\rho)} \, ,   
  \label{eq:XI_PI_P-}    
\end{eqnarray}   

 Thus classically all the variables can be rewritten in terms 
of those in the light-cone gauge. 
 Since we would like to calculate the commutation relations of 
super-Poincar\'e generators 
using the canonical commutation relations of those variables, 
we should check if these relations can be made 
into quantum mechanical relations, 
in such a way that the canonical commutation relations 
are satisfied. 
 What we are doing is basically a coordinate transformation 
$\rho \mapsto \sigma^2=\rho -A_{1\,{\rm lc}}(\rho)$. 
 Thus for quantities which do not involve the momentum $P_{A_1}$ 
conjugate to $A_{1\,{\rm lc}}(\rho)$, 
there is no ordering ambiguity, 
we can regard $A_{1\,{\rm lc}}(\rho)$ as a c-number. 
 The only variable about which we should be a little bit careful 
is $x^-_{\rm ml}(\sigma^2)$. 
 Since  $X^-_{\rm lc}$ on the right hand side 
of Eq.~(\ref{eq:X^-_0}) involves 
$P_{A_1}$ we should fix the ordering of the operators 
so that the canonical commutation relations are satisfied   
in the modified light-cone gauge.   
 Doing so is not difficult because $P_{A_1}$ appears only linearly   
in $X^-_{\rm lc}$.   
 By specifying the ordering so that we put $P_{A_1}$   
on the right of all the other operators,   
one can show that the commutation relations between   
operators involving $P_{A_1}$ at most linearly   
are the same as their classical counterparts.   
 Therefore what we should do is to check 
if the Poisson bracket 
\begin{equation} 
 [x^-_{\rm ml}(\sigma^2),\, p^+_{\rm ml}(\sigma^{\prime\,2})]_P 
 = \frac{1}{2\pi} \delta(\sigma^2 - \sigma^{\prime\,2}) \, , 
  \label{eq:X-P+} 
\end{equation} 
and others are satisfied classically or not. 
 Showing Eq.~(\ref{eq:X-P+}) is straightforward 
but needs some care. 
 Since $\sigma^2$ and $\rho$ are related to each other 
as Eq.~(\ref{eq:A_1}), 
$\rho$ can have nontrivial commutators 
with other variables, 
if we consider $\sigma^2$ as a c-number. 
 Anyway, from the commutator of the momentum constraints 
and other variables, one can show that the variables in  
the modified light-cone gauge 
satisfy the canonical commutation relations. 
  
 Thus we obtain the relations between 
the variables of the continuum membrane theory in two different gauges. 
 What we actually need is a discretized version of such relations. 
 Namely, we would like to obtain world sheet variables 
with independent $p^+,\,x^-$ 
in terms of the matrix string variables. 
 This can be done as follows. 
 Let us recall that the relations (\ref{eq:XI_PI_P-}) 
give a canonical transformation classically, 
and a unitary transformation in the quantum theory. 
 Therefore, we can get a unitary operator $\exp(i {\cal K})$ 
which translates the quantities at $(\sigma,\,\rho)$ 
in the light-cone gauge 
into the quantities at $(\sigma,\,\sigma^2)$ 
in the modified light-cone gauge. 
 Thus if we discretize the operator ${\cal K}$ in an appropriate way, 
by applying the unitary transformation corresponding to 
$\exp(i {\cal K})$ to the matrix string variables, 
we can get the variables corresponding to $N$ strings 
with independent $p^+,\,x^-$, 
which become those of modified light-cone gauge 
in the limit $N\rightarrow \infty$. 
  Thus, in principle, 
we can construct the world sheet variables of $N$ 
strings with independent $p^+,\,x^-$ 
from those of matrix string theory. 
 
 Wrapped membranes with nontrivial spatial topologies 
yield global constraints 
associated with closed but non-exact generators 
of area preserving diffeomorphism 
\cite{deWit:1997zq,deWit:1989vb,Uehara:2002vv}. 
 Since the DLCQ limit needs 
tentative compactification of the light-cone direction $x^-$, 
a global constraint shows up in solving $X^-$. 
 In this paper, our goal is a modest one, which is to realize 
membranes without any winding around $x^-$. 
 Of course, in order to construct the full-fledged membrane theory, 
we need to consider the sectors with nontrivial winding. 
 Since our main focus in this paper is the anomaly, 
it is sufficient to examine only the non-winding sector.

\subsection{super-Poincar\'e symmetry of matrix string theory} 
 With the explicit connection between two gauge choices, 
we can now see if there is super-Poincar\'e invariance 
at the leading order in matrix string theory 
by examining it in the modified light-cone gauge fixed membrane theory. 
 To make contact with the ultraviolet limit discussed 
in Subsec.~\ref{sec:remark}, 
we perform the rescaling which leads 
to the action (\ref{eq:lc_two_gauge}),  
together with the rescaled conjugate momenta $\Pi^I$ given by 
\begin{equation}   
 P^I = \frac{1}{R^{1/2}} \Pi^I \, .   
\end{equation} 
 The corresponding Hamiltonian density ${\cal H}$ 
is obtained by rescaling $(-P^-)$ determined from $\phi_0 = 0$ as 
\begin{equation} 
 - P^- = R {\cal H} \, . 
\end{equation} 
 The explicit form of ${\cal H}$ reads 
\begin{eqnarray} 
 {\cal H} &=&   
 \frac{1}{P^+}   
 \left(   
    \frac{1}{2}\,(\Pi^I)^2 + \frac{1}{2 R^3}\,(P_{A_1})^2   
  + \frac{1}{2}\,(\del_1 X^I)^2   
  + \frac{1}{4 R^3} (\{X^I,\,X^J\})^2   
 \right.   
  \nonumber \\ 
 && \qquad   
 \left.   
  + \frac{i}{2}\,\Psi^T \gamma^9 \del_1 \Psi 
  - i \frac{1}{R^{3/2}}\,\frac{1}{2}\,  
    \Psi^T \gamma^I \{ X^I,\,\Psi \}   
 \right) \, .   
  \label{eq:h_density}   
\end{eqnarray}   
 Let us define $\phi_{(m)}(\sigma^1)$ to be the value 
of a variable $\Phi(\sigma^1,\,\sigma^2 = ma)$   
at the lattice site $\sigma^2 = ma$   
($a \equiv \frac{2\pi}{N}$).   
 As anticipated, the Hamiltonian for $g_{\rm WM} = 0$   
turns out to be just 
the sum of   
the Hamiltonians of Green-Schwarz strings 
in the light-cone gauge;   
\begin{eqnarray}   
 H 
 &=&   
 a \sum_m \int_0^{2\pi} d \sigma^1\,{\cal H}   
  \nonumber \\ 
 &=&   
 a \sum_m  
   \int_0^{2\pi} d \sigma^1\,   
   \frac{1}{p^+_{(m)}} 
   \left(   
      \frac{1}{2} (\pi_{(m)}^I(\sigma^1))^2   
    + \frac{1}{2} (\del_1 x_{(m)}^I(\sigma^1))^2   
   \right.   
  \nonumber \\   
 && \qquad \qquad \qquad \qquad   
   \left.   
    + \frac{i}{2} s^a_{(m)}(\sigma^1) \del_1 s^a_{(m)}(\sigma^1)  
    - \frac{i}{2} 
       s^{\dot{a}}_{(m)}(\sigma^1) \del_1 s^{\dot{a}}_{(m)}(\sigma^1) 
   \right) \, ,  
\end{eqnarray}  
where $\Psi(\vec{\sigma})$ have been divided into 
two $8$ dimensional spinors, 
$S^a(\vec{\sigma}),\,S^{\dot{a}}(\vec{\sigma})$. 
 
 At least at the zeroth order in $\frac{1}{R^{3/2}}$,   
the super-Poincar\'e generators  
would be basically obtained   
as the sum of the super-Poincar\'e generators of strings 
put along the $\sigma^2$-direction. 
 The only thing one should take care is that since 
$x^+$ is common to all the strings, we take 
$p^-_{(m)}$ appearing in the generators to be the one 
obtained by solving the Hamiltonian constraint, 
except for the Fourier zero mode with respect to $m$. 
 For the zero mode, we use $p^-$ conjugate to $x^+$ as 
in the usual string theory.  
 With such a form, 
we can compute, say, 
$[L^{-I},\,L^{-J}]$ 
in the similar way as in the string case \cite{GSW} 
and find 
\begin{eqnarray} 
 \left[ L^{I-},\, L^{J-} \right] 
 &=& 
 a \sum_m 
 \frac{1}{(p^+_{(m)})^2} 
 \sum_{r=1}^\infty 
 \left( 
    \alpha^I_{(m)\,-r} \alpha^J_{(m)\,r} 
  - \alpha^J_{(m)\,-r} \alpha^I_{(m)\,r} 
 \right. 
  \nonumber \\ 
 && \qquad \qquad \qquad \quad  
 \left. 
  + \widetilde{\alpha}^I_{(m)\,-r} 
     \widetilde{\alpha}^J_{(m)\,r} 
  - \widetilde{\alpha}^J_{(m)\,-r} 
     \widetilde{\alpha}^I_{(m)\,r} 
 \right) 
  \nonumber \\ 
 && \qquad \qquad \qquad \qquad  
 \times 
 \frac{1}{a} \times 
 2 
 \left( 
  \frac{\delta^{KK}}{8} - 1 
 \right)\,r \, . 
\end{eqnarray} 
 Therefore 
the Schwinger term vanishes for $\delta^{KK} = 8$. 

 The calculation can be repeated for the supermembranes   
in the space-time with dimension equal to $D = 4,\,5$ or $7$,   
taking Majorana and/or Weyl properties into account.   
 In those cases, as expected, Lorentz anomaly exists.   
 Therefore the critical dimension for matrix string theory   
at the leading order is $D-1 = 10$. 

\section{Conclusions and discussions}   
\label{sec:concl}   
 In this paper, we have shown that   
wrapped membranes can be naturally   
related to matrix string theory via   
the rule proposed by Sekino and Yoneya,   
using the M(atrix) theory techniques.   
 From such a point of view, the membrane can be naturally   
regarded as a collection of strings interacting   
with each other.  
 In the lowest order in the perturbative expansion of   
matrix string theory, these strings become free. 
 With a change of variables discussed in Sec.~\ref{sec:Lorentz},   
the super-Poincar\'e generators of the theory can be   
constructed essentially as a sum of those of free strings   
and we can show that the theory is invariant under   
the super-Poincar\'e symmetry 
(except for the wrapped direction, $x^9$) 
if $D=10$ for matrix string theory 
or $D=11$ for the wrapped membrane.   
  
 Of course, this result is not sufficient to conclude   
that membrane theory is consistent only when $D=11$.   
 Since the matrix string action is that 
of a two dimensional gauge theory,   
the coupling constant is a dimensionful one.   
 Therefore in order to see if the theory is invariant under   
the super-Poincar\'e symmetry, we should check this for all   
orders in the perturbation theory.   
 It is quite challenging   
to see if we can construct the Lorentz generators   
at the next order of $\frac{1}{R^{3/2}}$.   
   
 We can also apply our approach   
to other types of membranes, e.g.,     
open membranes with the various types   
of boundary conditions.   
 The open membranes   
related to the heterotic strings in the context of M theory   
\cite{Horava_Witten}   
will be regularized by $O(N)$ matrices   
(See Ref.~\cite{Lowe:1997sx,Rey:1997hj,Horava:1997ns}).   
 In this context,   
the membrane may be regarded as being composed   
of the interacting strings put at the sites   
along the spatial direction with boundary.   
 It is interesting to see 
if the consistency conditions of the   
theory require the existence of   
appropriate number of fermions on each boundary 
which should provide $E_8$ gauge bosons. 

\acknowledgments 
 The authors would like to thank the colleagues in Theory Group at KEK 
for discussions, 
and for providing opportunities and space to have discussions 
after leaving there. 
 The works of both authors are supported in part 
by Grant-in-Aid for Scientific Research from 
the Ministry of Education, Science and Culture in Japan. 
  
\section*{Appendix}   
 This appendix summarizes   
the convention for t'Hooft matrices   
and the related objects used in the text.   
 We assume that the rank $N$ of the matrices treated is even   
\footnote{   
 See, for instance, Ref.~\cite{FFZ}.   
}.   
 Using   
$\omega_{(N)} \equiv \exp\left(i \frac{2\pi}{N} \right)$,   
$N \times N$ t'Hooft matrices $U_{(N)}$, $V_{(N)}$   
satisfy the relation   
\begin{equation}   
 U_{(N)} V_{(N)} = \omega_{(N)} V_{(N)} U_{(N)} \, .   
\end{equation}   
 One explicit representation of such $U_{(N)}$, $V_{(N)}$ is   
\begin{eqnarray}   
 U_{(N)}   
 &=& e^{iP_{(N)}}   
 =  \left(   
     \begin{array}{ccccccccc}   
      0 & 1 & 0 & \cdot & \cdot & \cdot & 0 & 0 & 0\\   
      0 & 0 & 1 & \cdot & \cdot & \cdot & 0 & 0 & 0\\   
      0 & 0 & 0 & \cdot & \cdot & \cdot & 0 & 0 & 0\\   
        &   &   & \cdot & \cdot &       &   &   & \\   
        &   &   &       & \cdot & \cdot &   &   & \\   
        &   &   &       &       & \cdot & \cdot & & \\   
        &   &   &       &       &       & \cdot & \cdot & \\   
        &   &   &       &       &       &       & \cdot & 1\\   
      1 & 0 & 0 & \cdot & \cdot & \cdot & 0 & 0 & 0   
     \end{array}   
    \right) \, ,   
 \nonumber \\   
 V_{(N)}   
 &=& e^{iQ_{(N)}}   
 = \left(   
     \begin{array}{cccccc}   
      1 & & & & & \\   
      & \omega_{(N)} & & & & \\   
      & & \cdot & & & \\   
      & & & \cdot & & \\   
      & & & & \cdot & \\   
      & & & & & \left( \omega_{(N)} \right)^{N-1}   
     \end{array}   
    \right) \, .   
 \label{eq:repr_UV}   
\end{eqnarray}   
 They satisfy $(U_{(N)})^N = 1 = (V_{(N)})^N$.   
   
 A basis of of Lie algebra of $U(N)$ is given by   
$\left\{ J^{(N)}_{\vec{m}} \right\}_{\vec{m}}$,   
where $\vec{m} \equiv (m_1, m_2)$ is a set of two integers $m_1, m_2$   
and   
\begin{equation}   
 J^{(N)}_{\vec{m}}   
  \equiv \left( \omega_{(N)} \right)^{-\frac{1}{2} m_1 m_2}   
         \left( U_{(N)} \right)^{m_1}   
         \left( V_{(N)} \right)^{m_2}   
  \, .   
   \label{eq:J}   
\end{equation}   
 One can check that   
$(J^{(N)}_{\vec{m}})^\dagger = J^{(N)}_{-\vec{m}}$,   
and   
\begin{eqnarray}   
 &   
 \displaystyle{   
  J^{(N)}_{\vec{m}} J^{(N)}_{\vec{n}}   
  = \left( \omega_{(N)} \right )^{\frac{1}{2} \vec{m} \times \vec{n}}   
    J^{(N)}_{\vec{m} + \vec{n}} \, ,   
 }& \nonumber \\   
 &   
 \displaystyle{   
  \left[ J^{(N)}_{\vec{m}}, J^{(N)}_{\vec{n}} \right]   
  = 2i   
    \sin\left( \frac{\pi}{N} \vec{m} \times \vec{n} \right)   
    J^{(N)}_{\vec{m} + \vec{n}} \, , 
 }&   
\end{eqnarray}   
with $\vec{m} \times \vec{n} \equiv m_1 n_2 - m_2 n_1$. 
  
   

\begin{thebibliography}{99}   
 \bibitem{matrix_reg}   
  J.~Goldstone, unpublished; \\   
  J. Hoppe,   
   Proc.~Int.~Workshop on   
   {\it Constraints Theory and Relativistic Dynamics},   
   edited by G.~Longhi and L.~Lusannna,   
   World Scientific (Singapore), 1987.   
%
\bibitem{deWit:1988ig}   
 B.~de Wit, J.~Hoppe and H.~Nicolai,  
  {\it On the quantum mechanics of supermembranes},   
  Nucl.\ Phys.\ B {\bf 305} (1988) 545. 
%
\bibitem{Banks:1996vh}   
 T.~Banks, W.~Fischler, S.~H.~Shenker and L.~Susskind,   
  {\it M theory as a matrix model: A conjecture},   
  Phys.\ Rev.\ D {\bf 55} (1997) 5112,   
  hep-th/9610043. 
%
\bibitem{Sen:1997we}   
 A.~Sen,  
  {\it D0 branes on $T^n$ and matrix theory},   
  Adv.\ Theor.\ Math.\ Phys.\  {\bf 2} (1998) 51,   
  hep-th/9709220. 
%
\bibitem{Seiberg:1997ad}   
 N.~Seiberg,   
  {\it Why is the matrix model correct?},   
  Phys.\ Rev.\ Lett.\  {\bf 79} (1997) 3577,   
  hep-th/9710009. 
%
 \bibitem{M_theory}   
  M.~J.~Duff, P.~S.~Howe, T.~Inami and K.~S.~Stelle,  
   {\it Superstrings in D = 10 from supermembranes in $D = 11$},   
   Phys.\ Lett.\ B {\bf 191} (1987) 70; \\ 
  C.~M.~Hull and P.~K.~Townsend,   
   {\it Unity of superstring dualities},   
   Nucl.\ Phys.\ B {\bf 438} (1995) 109,   
   hep-th/9410167; \\  
  P.~K.~Townsend,   
   {\it The eleven-dimensional supermembrane revisited},   
   Phys.\ Lett.\ B {\bf 350} (1995) 184,    
   hep-th/9501068; \\ 
  E.~Witten,  
   {\it String theory dynamics in various dimensions},   
   Nucl.\ Phys.\ B {\bf 443} (1995) 85,   
   hep-th/9503124. 
%
\bibitem{deWit:1997zq} 
 B.~de Wit, K.~Peeters and J.~Plefka,
  {\it Supermembranes with winding}, 
  Phys.\ Lett.\ B {\bf 409} (1997) 117, 
  arXiv:hep-th/9705225; \\ 
  {\it The supermembrane with winding}, 
  Nucl.\ Phys.\ Proc.\ Suppl.\  {\bf 62} (1998) 405, 
 arXiv:hep-th/9707261. 
%
\bibitem{Sekino:2001ai}  
 Y.~Sekino and T.~Yoneya,  
  {\it From supermembrane to matrix string},   
  Nucl.\ Phys.\ B {\bf 619} (2001) 22,   
  hep-th/0108176. 
%
\bibitem{matrix_string}   
 L.~Motl,   
  {\it Proposals on nonperturbative superstring interactions},   
  hep-th/9701025; 
  \\   
 T.~Banks and N.~Seiberg,  
  {\it Strings from matrices},   
  Nucl.\ Phys.\ B {\bf 497} (1997) 41,   
  hep-th/9702187; 
  \\ 
 R.~Dijkgraaf, E.~Verlinde and H.~Verlinde,   
  {\it Matrix string theory},   
  Nucl.\ Phys.\ B {\bf 500} (1997) 43,   
  hep-th/9703030. 
%
\bibitem{Uehara:2002ga}   
 S.~Uehara and S.~Yamada,   
  {\it On the strong coupling region in quantum matrix string theory},   
  JHEP {\bf 0209} (2002) 019,   
  hep-th/0207209. 
%
\bibitem{Cederwall:2002wh}   
 M.~Cederwall,  
  {\it Open and winding membranes,   
  affine matrix theory and matrix string theory}, 
  JHEP {\bf 0212} (2002) 005, 
  hep-th/0210152. 
\bibitem{Shimada:2003ks} 
 H.~Shimada,
  {\it Membrane topology and matrix regularization}, 
  hep-th/0307058. 
\bibitem{Recent_membrane} 
 H.~Sugawara, 
  {\it Theory of membrane in Heegaard diagram expansion}, 
  hep-th/0304164; \\ 
 C.~Zachos,
  {\it Membranes and consistent quantization of Nambu dynamics}, 
  Phys.\ Lett.\ B {\bf 570} (2003) 82, 
  hep-th/0306222; \\ 
 J.~Dai and Y.~S.~Wu,
  {\it Quiver matrix mechanics for IIB string theory. I 
   : Wrapping membranes and emergent dimension}, 
  hep-th/0312028. 
\bibitem{Taylor:1996ik}   
 W.~I.~Taylor,  
  {\it D-brane field theory on compact spaces},   
  Phys.\ Lett.\ B {\bf 394} (1997) 283,   
  hep-th/9611042. 
%
\bibitem{Giddings:1998yd}   
 S.~B.~Giddings, F.~Hacquebord and H.~Verlinde,   
  {\it High energy scattering and D-pair creation in matrix string theory},   
  Nucl.\ Phys.\ B {\bf 537} (1999) 260,   
  hep-th/9804121. 
%
 \bibitem{HI}  
  M.~Hayakawa and N.~Ishibashi,  
   {\it Perturbative world-volume dynamics   
    of the bosonic membrane and string},   
   Nucl.\ Phys.\ B {\bf 614} (2001) 171,   
   hep-th/0107103. 
%
\bibitem{Bergshoeff:1987cm}   
 E.~Bergshoeff, E.~Sezgin and P.~K.~Townsend,   
  {\it Supermembranes and eleven-dimensional supergravity},   
  Phys.\ Lett.\ B {\bf 189} (1987) 75; 
  \\ 
  {\it Properties of the eleven-dimensional super membrane theory},   
  Annals Phys.\  {\bf 185} (1988) 330. 
%
 \bibitem{B_membrane}    
  U.~Marquard and M.~Scholl,  
   {\it Conditions on the embedding space of p-branes   
    from their constraint algebras},   
   Phys.\ Lett.\ B {\bf 209} (1988) 434; 
   \\   
   {\it Lorentz algebra and critical dimension   
    for the bosonic membrane},   
   Phys.\ Lett.\ B {\bf 227} (1989) 227; 
   \\   
  I.~Bars,   
   {\it Membrane symmetries and anomalies},   
   Nucl.\ Phys.\ B {\bf 343} (1990) 398. 
%
 \bibitem{S_membrane}   
  I.~Bars, C.~N.~Pope and E.~Sezgin,  
   {\it Massless spectrum and critical dimension of the supermembrane},   
   Phys.\ Lett.\ B {\bf 198} (1987) 455; 
   \\    
  U.~Marquard, R.~Kaiser and M.~Scholl,   
   {\it Lorentz algebra and critical dimension for the supermembrane},   
   Phys.\ Lett.\ B {\bf 227} (1989) 234. 
%
\bibitem{deWit:1989vb}
 B.~de Wit, U.~Marquard and H.~Nicolai,  
  {\it Area-preserving diffeomorphisms and supermembrane 
   Lorentz invariance},   
  Commun.\ Math.\ Phys.\  {\bf 128} (1990) 39. 
%
\bibitem{LI_matrix} 
 K.~Ezawa, Y.~Matsuo and K.~Murakami,   
  {\it Lorentz symmetry of supermembrane in light cone gauge formulation},   
  Prog.\ Theor.\ Phys.\  {\bf 98} (1997) 485,    
  hep-th/9705005; 
  \\   
 D.~A.~Lowe,  
  {\it Eleven-dimensional Lorentz symmetry from SUSY quantum mechanics},   
  JHEP {\bf 9810} (1998) 003,   
  hep-th/9807229. 
%
\bibitem{Uehara:2002vv} 
 S.~Uehara and S.~Yamada, 
  {\it Comments on the global constraints in light-cone string 
  and membrane theories}, 
  JHEP {\bf 0212} (2002) 041, 
  arXiv:hep-th/0212048.
%
\bibitem{GSW}   
 M.~B.~Green, J.~H.~Schwarz and E.~Witten,   
 {\it Superstring Theory. Vol. 1: Introduction}, 
 (Cambridge University Press, 1987).  
%
\bibitem{Horava_Witten}  
 P.~Horava and E.~Witten,  
  {\it Heterotic and type I string dynamics from eleven dimensions},   
  Nucl.\ Phys.\ B {\bf 460} (1996) 506,   
  hep-th/9510209; 
  \\   
  {\it Eleven-Dimensional Supergravity on a Manifold with Boundary},   
  Nucl.\ Phys.\ B {\bf 475} (1996) 94,   
  hep-th/9603142. 
%
\bibitem{Lowe:1997sx} 
 D.~A.~Lowe, 
  {\it Heterotic matrix string theory},   
  Phys.\ Lett.\ B {\bf 403} (1997) 243,   
  hep-th/9704041; 
%
\bibitem{Rey:1997hj}
 S.~J.~Rey,  
  {\it Heterotic M(atrix) strings and their interactions},   
  Nucl.\ Phys.\ B {\bf 502} (1997) 170,  
  hep-th/9704158; 
%
\bibitem{Horava:1997ns}
 P.~Horava,  
  {\it Matrix theory and heterotic strings on tori},   
  Nucl.\ Phys.\ B {\bf 505} (1997) 84,  
  hep-th/9705055. 

\bibitem{FFZ}   
 D.~B.~Fairlie and C.~K.~Zachos,  
  {\it Infinite dimensional algebras, sine brackets and $SU(\infty)$},   
  Phys.\ Lett.\ B {\bf 224} (1989) 101; 
  \\   
 D.~B.~Fairlie, P.~Fletcher and C.~K.~Zachos,  
  {\it Infinite dimensional algebras and a trigonometric basis   
  for the classical Lie algebras},   
  J.\ Math.\ Phys.\  {\bf 31} (1990) 1088. 
\end{thebibliography}
\end{document}